# LISTEN: Lightweight Industrial Sound-representable Transformer for Edge Notification

Changheon Han[1], Yun Seok Kang[2], Yuseop Sim[1], Hyung Wook Park[2], and Martin Byung-Guk Jun[1*]

[1]School of Mechanical Engineering, Purdue University, 585 Purdue Mall, West Lafayette 47907, USA.

[2]Department of Mechanical Engineering, UNIST, Unist-gil 50, Eonyang-eup, Ulju-gun, Ulsan, Korea 44919

*Corresponding author(s). E-mail(s): mbgjun@purdue.edu

## Abstract

Deep learning-based machine listening is broadening the scope of industrial acoustic analysis, yet its widespread implementation on live shop floors is hindered by the reliance on large, task-specific annotated datasets for every new task. While emerging general-purpose sound foundation models aim to alleviate data dependency, they reveal critical dilemmas in practice. General-purpose sound foundation models are computationally expensive and fail in industrial scenarios characterized by tonal harmonics, broadband noise, and transient fault events, making instant, on-site deployment impractical. These challenges combined mean that a practical, end-to-end system for deploying a sound foundation model on a live shop floor has remained elusive. To address this challenge, this study introduces LISTEN (Lightweight Industrial Sound-representable Transformer for Edge Notification), the first lightweight foundation model specialized for industrial sound. Through Knowledge Distillation (KD) from the large-scale teacher model IMPACT (Industrial Machine Perception via Acoustic Cognitive Transformer), we construct LISTEN optimized for resource-constrained edge environments. By freezing the backbone and training only a shallow head on minimal target-process data, rather than performing full fine-tuning or retraining, LISTEN achieves nearly identical performance to IMPACT across diverse manufacturing processes. This study further demonstrates a complete system for real-time machine monitoring, encompassing data acquisition with Industrial Internet of Things (IIoT) devices, rapid model adaptation using minimal annotated data, and real-time monitoring on a low-cost edge device. By validating the entire system on a live Computerized Numerical Control (CNC) machine, this work establishes the first feasible end-to-end system for deploying a lightweight industrial sound foundation model in an active industrial environment.

*The codes and pre-trained models are available at https://github.com/hanprd/LISTEN*

## 1. Introduction

Audio from industrial machinery serves as a critical source for anomaly detection, predictive maintenance, and process optimization—essential tasks for improving reliability and efficiency in manufacturing [1]. Yet, the current deep learning approaches for machine listening are typically supervised and task-specific. Their reliance on large, manually annotated datasets for every new task makes it difficult and costly to generalize solutions across diverse industrial scenarios.

Having proven effective in Natural Language Processing (NLP) [2,3] and computer vision [4–6], foundation models that learn general-purpose representations from vast unlabeled data offer a path around this data bottleneck. Most sound foundation models [7–9], however, are trained on general sounds like speech or music. Industrial acoustics are fundamentally different, with distinct signatures such as tonal harmonics from rotating parts, broadband noise from friction, and transient events that signal equipment faults [10].

Furthermore, the mechanisms of industrial sound generation vary across machines. For example, Laser Powder Bed Fusion (LPBF) systems primarily produce sound from mechanical subsystem motion, whereas Computerized Numerical Control (CNC) machines involve spindle rotation and tool-workpiece interactions, resulting in parameter-dependent harmonic structures with fine spectral–temporal variations. In contrast, cold spray equipment is governed by high-pressure gas–powder delivery dynamics and feeder-state interactions. These differences require models to capture both global and local acoustic characteristics across diverse machine types. As a result, models trained on general audio often fail to capture these critical features.

To address these challenges, recent work has introduced two key resources: DINOS (Diverse INdustrial Operation Sounds) [11], a large-scale, open-access dataset with over 1,000 hours of diverse industrial processes, and IMPACT (Industrial Machine Perception via Acoustic Cognitive Transformer) [12], the first industrial sound foundation model. IMPACT is trained on DINOS using the Efficient Audio Transformer (EAT) [13], a self-supervised learning architecture designed to capture both local and global spectrogram features. LISTEN is developed through Knowledge Distillation (KD) from this domain-specific teacher model, inheriting its ability to effectively represent both global and local characteristics of industrial sound. When adapted to over 30 distinct downstream industrial acoustic tasks, the resulting model outperforms general-purpose sound foundation models such as CLAP, VGGish, and AudioMAE. This result highlights the advantages of a domain-specific approach for industrial scenarios.

However, foundation models, including IMPACT, often suffer from a major flaw: they are too large. Their size demands cloud systems or powerful GPUs for adaptation and inference, making them impractical and too expensive for widespread, real-time monitoring on low-cost edge devices used on shop floors. This critical bottleneck has prevented the development of a practical, end-to-end system for deploying a sound foundation model on a live shop floor, despite recent advances in model performance.

To fill this gap, this study introduces a complete, end-to-end system for on-device industrial machine monitoring, with its workflow illustrated in Fig. 1. At the core of this system is LISTEN (Lightweight Industrial Sound-representable Transformer for Edge Notification), a novel foundation model for industrial sound with a size of only a few hundred kilobytes. First, as shown in (a) Model Lightweighting, we create an optimized teacher model, O'IMPACT, by applying a grid search to the IMPACT model. We then distill the LISTEN model from O'IMPACT. The model's comprehensive performance is evaluated on 30 distinct downstream tasks from the DINOS dataset. In this evaluation, a shallow Multilayer Perceptron (MLP) head is trained on downstream datasets while freezing each backbone.

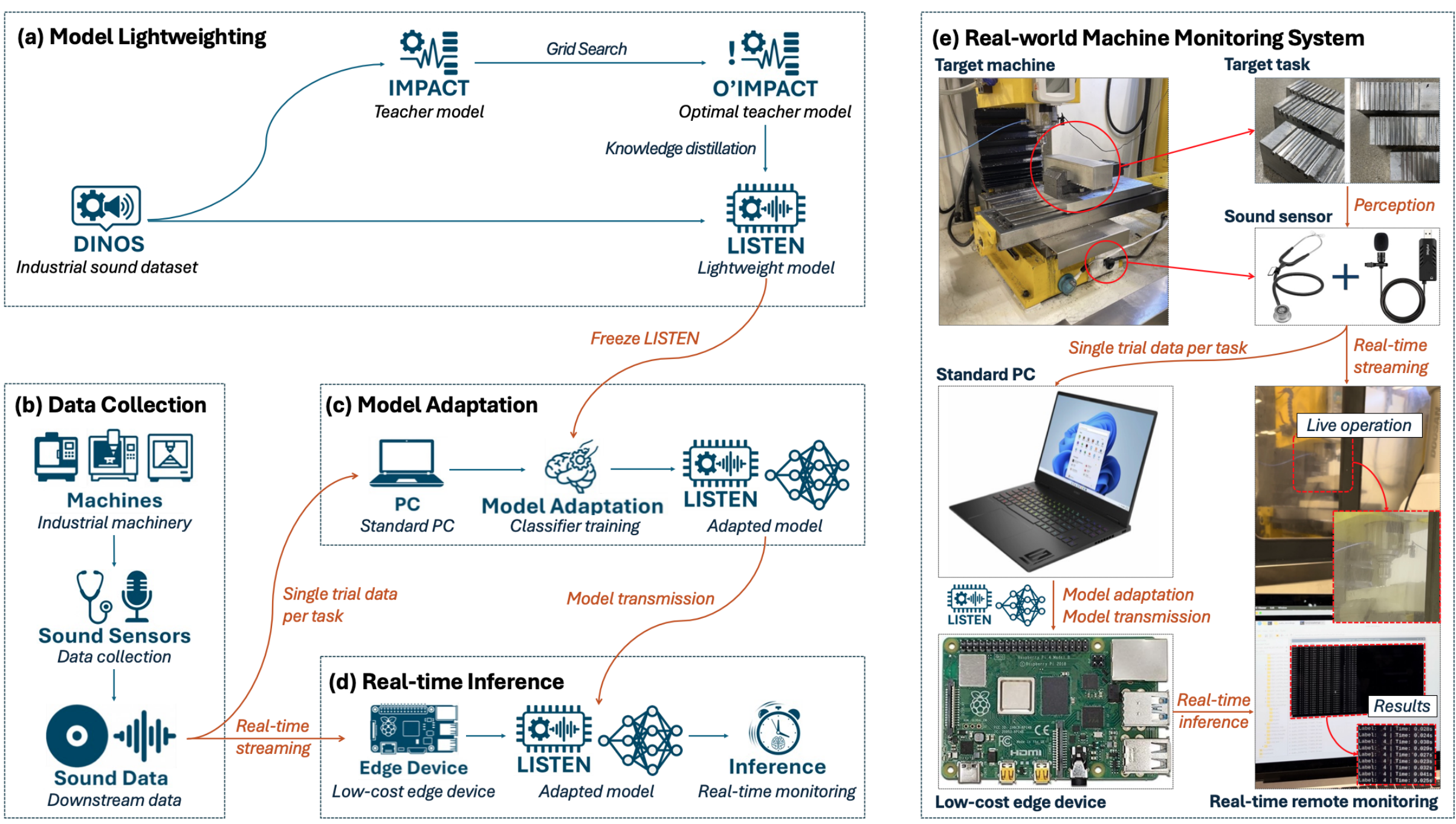

**Fig. 1** End-to-End Workflow of the LISTEN System

Second, to enable its practical utility in a new environment, we propose a novel on-site deployment workflow, illustrated in Fig. 1(b-d). This workflow is designed for rapid adaptation using minimal data. It begins with (b) Data Collection, where a small dataset is captured from a new target machine, often from a single trial. This data is then used for (c) Rapid Model Adaptation, which trains a shallow MLP head on a standard PC with the LISTEN backbone kept frozen. The resulting specialized model is subsequently deployed for (d) Real-time Inference on a low-cost edge device, such as a Raspberry Pi.

Finally, we validate this entire proposed workflow through a real-world implementation. As depicted in Fig. 1(e) Real-world Machine Monitoring System, these phases integrate the preceding steps into a complete and practical

Industrial Internet of Things (IIoT) framework on a live shop floor. This system demonstrates the full cycle from live machine sensing to on-device inference and remote monitoring, establishing a feasible pathway for deploying advanced Artificial Intelligence (AI) on active industrial environments using low-cost edge devices.

Here are the summarized contributions of this work:

- **The first lightweight foundation model for industrial sound.** We introduce LISTEN, a highly compact and efficient model specifically designed for the acoustic characteristics of industrial environments. The grid search and distillation process is performed only once during the lightweighting stage. For a new dataset or environment, LISTEN does not require full fine-tuning or retraining. Instead, it supports rapid adaptation by training only a shallow MLP head with minimal data. Its development directly addresses the critical limitations of model size and computational cost that have prevented the use of foundation models on shop-floor edge devices.
- **A complete, end-to-end system for on-device machine monitoring.** We propose a practical workflow that encompasses the entire lifecycle of the model. This includes the initial KD to create LISTEN, a rapid on-site model adaptation process using minimal live data without full fine-tuning or retraining, and final deployment for real-time inference on low-cost hardware. Each stage is designed to reduce overall system complexity during deployment. KD enables compressing a large-scale teacher model with validated generalizable industrial sound representations into a lightweight architecture. This allows deployment to rely only on lightweight adaptation of a shallow head. This decoupling between model development and on-site deployment eliminates the need for repeated large-scale training when new data or environments are introduced, thereby significantly simplifying real-world implementation.
- **An initial real-world validation of the lightweight industrial sound foundation model.** Initially, the scalability and generalizability of LISTEN are systematically validated on the DINOS benchmark across unseen physical equipment, machine models, and processes. Subsequently, the feasibility and effectiveness of LISTEN are demonstrated in an active industrial environment through full deployment on a live CNC machine. In this setting, LISTEN promptly adapts to a new task using only minimal data, while maintaining competitive performance. This combined validation confirms the practical utility of LISTEN and establishes a crucial reference and a clear pathway toward the practical application of advanced acoustic condition monitoring on live shop floors.

This paper is organized as follows. Section 2 reviews related works. Section 3 details the methodology of this study, describing the development of LISTEN and its implementation on the shop floor. Section 4 presents experimental results and analysis. Section 5 concludes the paper with a summary and directions for future research.

# 2. Related Works

## 2.1. From Signal Processing to Deep Learning in Acoustic Condition Monitoring

Acoustic condition monitoring for industrial machinery has evolved significantly from classical signal processing methods to recent advanced data-driven approaches. Traditionally, Machine Condition Monitoring (MCM) has primarily relied on vibration analysis, utilizing contact-based sensors to capture vibration signatures. Mechanical faults like bearing wear or gear damage generate characteristic frequencies, detectable through spectral and envelope analysis [10]. Acoustic condition monitoring offers a contactless alternative, leveraging the principle that physical phenomena causing vibration also generate sound waves [14]. This eliminates the need for direct sensor mounting, which can be difficult or hazardous in industrial environments. However, high ambient noise and complex sound propagation paths often degrade the signal-to-noise ratio, obscuring fault-related acoustic signatures.

Early acoustic MCM systems addressed these challenges with feature engineering techniques to extract meaningful information from raw audio signals. Mel-Frequency Cepstral Coefficients (MFCCs) emerged as a highly successful and widely used feature set. Based on their proven effectiveness in speech recognition by Davis and Mermelstein [15], MFCCs offer a compact representation of the spectral characteristics of sound. They are designed to mimic the way the human auditory system perceives sound frequencies, making them effective at capturing relevant acoustic information. MFCCs and similar handcrafted features have been the standard input for machine learning classifiers in fault detection systems [16–18].

The advent of deep learning has transformed acoustic MCM, shifting from manual feature engineering to automated representation learning. Marchi et al. [19] revealed that deep recurrent autoencoders could effectively identify deviations from normal machine operation by measuring reconstruction errors. This is especially useful for industrial monitoring due to the inherent class imbalance in the available data; normal operational data is abundant, while fault data is rare and diverse. This imbalance makes traditional supervised learning impractical, and hence unsupervised anomaly detection has become a promising approach. Autoencoders, which learn to reconstruct their input data, proved to be a cornerstone technology for this purpose [20].

The transition from handcrafted features to learned representations marks a fundamental move towards fully data-driven methods, paving the way for even more advanced deep learning architectures in acoustic condition monitoring.

## 2.2. Transformer Paradigm and Rise of Foundation Models

Introduced by Vaswani et al. [21], the transformer architecture has triggered a major paradigm shift in AI. Originally developed for NLP, its principles have been adapted for numerous other domains, enabling the rise of large-scale, pre-trained foundation models. As a result, transformers have replaced the Recurrent and Convolutional Neural Networks (RNNs and CNNs) that were conventionally used for sequence modeling tasks.

The key innovation of the transformer architecture is the self-attention mechanism. This allows the model to weigh the importance of every element in an input sequence relative to all other elements, enabling it to capture global dependencies regardless of their distance in the sequence. A major advantage of this design is its high parallelizability. Unlike sequential architectures like RNNs, transformers can process all elements of a sequence simultaneously, which dramatically accelerates training on modern hardware like GPUs. These features have spurred the development of large-scale foundation models, which are pre-trained on massive datasets and can discern subtle patterns in complex scenarios.

Foundation models typically follow a two-stage workflow: pre-training and model adaptation. In the first stage, a large transformer model is trained on vast amounts of unlabeled data by self-supervised or unsupervised learning. For instance, BERT [2] was pre-trained to predict masked tokens from surrounding context, while GPT [22] models learned to predict the next token in a sequence. This process imbues the model with general-purpose representations. In the second stage, the pre-trained model is adapted on a smaller task-specific labeled dataset to achieve high performance on a particular downstream application.

The success of transformers in NLP inspired adaptation to other modalities. The Vision Transformer (ViT) showed that transformers could achieve State-of-the-Art (SOTA) results in image classification by treating an image as a sequence of patches [23]. In this setup, an image is divided into fixed-size patches, which are then flattened, and fed them into transformers as a sequence. This approach proves their flexibility in handling spatial data.

This idea was soon carried into audio with the introduction of the Audio Spectrogram Transformer (AST) [24]. The AST converts a raw audio waveform into a two-dimensional log-mel spectrogram and processes it identically to how ViT processes an image. By treating the spectrogram as an image, the AST became the first fully attention-based model for audio classification to achieve SOTA performance across multiple benchmarks. This pragmatic approach of leveraging mature vision architectures for audio data reflects the current trend in the field and forms the foundation for our own work.

More recently, the EAT [13] introduced a SOTA self-supervised learning framework designed to capture more comprehensive audio representations. EAT employs a student-teacher architecture where two identical transformer models are used. During training, the student model learns from a spectrogram with 70% of its patches masked, while the teacher model processes the complete, unmasked spectrogram. The framework is optimized using a dual-objective loss function that combines a local loss and a global loss, compelling the model to learn both fine grained details and broader temporal structures within audio.

## 2.3. State-of-the-Art in Sound Foundation Models

Recent research has increasingly focused on developing large-scale pre-trained sound models through self-supervised or multimodal learning, with a growing trend towards domain specialization. Masked autoencoding on spectrograms has emerged as a powerful self-supervised learning strategy. For example, AudioMAE is pre-trained by masking a substantial portion of input spectrogram patches and training the model to reconstruct the original form from the visible patches [25]. This encourages the model to learn meaningful structural representations of audio structure without relying on labeled data.

An alternative line of work leverages multimodal learning. CLAP [26] trains audio and text encoders simultaneously on paired audio clips and natural language descriptions, aligning their representations in a shared embedding space. This design allows for zero-shot classification, enabling the model to generalize to new tasks without additional training. While general-purpose models like AudioMAE and CLAP show broad utility across a range of tasks, there is growing evidence that domain-specific pre-trained models yield superior performance on specialized tasks.

A compelling example is OPERA [27], a model pre-trained on respiratory sounds for medical applications, which outperformed generalist audio models on 16 out of 19 downstream respiratory health tasks. This precedent strongly suggests that foundation models achieve superior performance when specialized for the acoustic characteristics of a target domain. This finding was recently corroborated in the industrial sound domain. Han et al. [12] developed the first industrial sound foundation model (IMPACT) by training a SOTA architecture (EAT) on the domain-specific dataset (DINOS). The effectiveness of this approach was demonstrated as IMPACT outperformed general-purpose models, such as VGGish, CLAP, and AudioMAE, on 24 of 30 distinct industrial downstream tasks.

## 2.4. Imperative for Lightweight Models in Industrial Internet of Things

While the industrial foundation models, IMPACT, are powerful, their computational cost is a significant barrier to deployment in practical industrial settings. Modern industrial systems rely on IIoT, where real-time data from connected sensors informs operational decisions. For time-sensitive applications like fault detection, sending raw sensor data to a central server is often impractical due to latency and bandwidth constraints while also posing significant cybersecurity risks. This necessitates edge computing, where model inference is performed on resource-constrained local devices, such as Raspberry Pi. A primary obstacle to deploying standard foundation models on edge devices is the self-attention mechanism, whose computational and memory complexity scales quadratically with the input sequence length [28]. For high-resolution audio, this quadratic scaling is often computationally prohibitive on edge hardware.

To mitigate this bottleneck, researchers have developed more efficient transformer architectures. Hybrid models like MobileViT combine efficient convolutional layers for local feature extraction with a compact transformer for global representation learning [29]. Other approaches directly optimize the attention mechanism itself [30]. Designing new, efficient architectures from scratch is one possible approach. However, this process often requires navigating a vast search space for hyperparameter optimization, which significantly increases training difficulty and computational cost [31]. Furthermore, this approach fails to leverage the rich, nuanced knowledge already captured by an existing large-scale model.

Instead, techniques that improve computational efficiency while retaining the performance of existing models can be considered as alternatives for resource-constrained devices [32]. First, quantization compresses model size by reducing the numerical precision of parameters. Chen et al. [33] proposed Efficient Quantization-Aware Training (EfficientQAT), which optimizes a Large Language Model (LLM) into a 2-bit model with less than 3 points of accuracy degradation. Nevertheless, quantization has inherent limitations. The maximum compression ratio of quantization is bound by the bit-width ratio between the original and the reduced precision formats. Moreover, the lack of native hardware support can result in computational overhead due to the dequantization process required to convert low-precision data back into FP32 [34,35].

Next, pruning reduces model size and computational complexity by eliminating redundant or less important weights [36]. This technique is generally categorized into unstructured and structured pruning. Unstructured pruning targets individual weights based on magnitude or importance criteria, resulting in highly sparse models. Although it may achieve high compression ratios, realizing actual inference speedup often requires specialized hardware or software support due to its sparsity patterns. In contrast, structured pruning removes entire parameter groups, enabling hardware-agnostic inference acceleration. However, this coarse-grained removal necessitates careful parameter fine-tuning or retraining to prevent performance degradation caused by the loss of critical model components.

Alternatively, KD is a strategy that addresses this limitation. The primary advantage of KD is its ability to transfer complex relationships between data points, often referred to as dark knowledge, from a large, pre-trained teacher model to a much smaller and more efficient student model [37]. Given that a domain-specific teacher model (IMPACT) was available for this study, KD presented the most logical and efficient pathway for creating a high-performance lightweight model. KD has been successfully applied to compress large audio models for deployment on low-resource devices [38].

KD strategies are generally categorized into feature-based, relation-based, and response-based methods [39]. Feature-based distillation aims to capture comprehensive representations by mapping intermediate features. However, it often faces semantic mismatch problems due to capacity gaps or architectural heterogeneity between teacher and student models, requiring auxiliary projection layers. Relation-based distillation, while effective in capturing high-order relationships across samples and layers, is computationally expensive with complexity scaling quadratically with batch size and vulnerable to performance degradation due to unstable relation definitions in ambiguous scenarios [40]. Conversely, response-based distillation [41] focuses on outcome-driven learning by mimicking the teacher's final prediction logits. This approach offers architectural flexibility, allowing the student to be designed independently of the teacher's internal structure. Such independence is advantageous to develop extremely lightweight models for low-cost edge devices.

## 2.5. Synthesis and Identified Research Gap

Our review of the literature reveals a clear dichotomy. On one hand, large-scale foundation models have demonstrated that performance on industrial tasks is maximized through domain-specific pre-training, as shown by models like IMPACT. On the other hand, a separate body of work has produced various techniques for creating lightweight models suitable for edge computing, such as efficient architectures and knowledge distillation.

However, these two streams of research have not yet effectively converged for industrial acoustics. Powerful, domain-specific models like IMPACT remain too computationally expensive for practical on-site deployment. Conversely, existing lightweight models are typically general-purpose and lack the specialized pre-training required for robust performance in complex industrial environments. This leads to a critical and well-defined research gap. Specifically, there is an absence of a foundation model that is both pre-trained for the unique characteristics of industrial machinery and architecturally optimized for real-time inference on low-cost edge devices.

To bridge this gap, our study first leverages the advantages of KD to create LISTEN that retains IMPACT's domain-specific expertise within a highly efficient architecture. The primary contribution of this work, however, is not the model alone but the proposal and validation of a complete framework that enables practical, on-device intelligent monitoring in live industrial environments. The following section presents the detailed methodology.

# 3. Methodology

The methodology of this study is organized into two main phases. The first phase details the development of LISTEN through KD. To achieve this, we first identify the optimal hyperparameters for the parent IMPACT model, creating an optimized version named O'IMPACT. We then distill the knowledge from O'IMPACT into LISTEN's compact architecture, using a grid search to find the optimal configuration while evaluating its performance and inference time directly on a target edge device (Raspberry Pi 4 8GB). The second phase demonstrates the practical validation of our proposed end-to-end system. In this phase, we implement LISTEN on the edge device to monitor a new CNC machining scenario, verifying the entire workflow from rapid on-site model adaptation to real-time inference in a live industrial environment.

## 3.1. Model Development and Evaluation Protocol

Fig. 2 illustrates the KD pipeline to develop LISTEN. In this process, LISTEN was optimized by minimizing the Mean Squared Error (MSE) loss relative to a teacher model. To identify the optimal teacher model, we conducted a grid search by varying key transformer hyperparameters. Our grid search focused on the embedding dimensions and the number of layers, as it is well-established that these two hyperparameters, often referred to as model width and depth respectively, are the most dominant factors that determine a transformer's parameter count and representational capacity [21,42]. Table 1 lists the hyperparameters evaluated during this grid search on the IMPACT architecture, while all other transformer, CNN encoder, and CNN decoder hyperparameters were kept the same as those of the original IMPACT model.

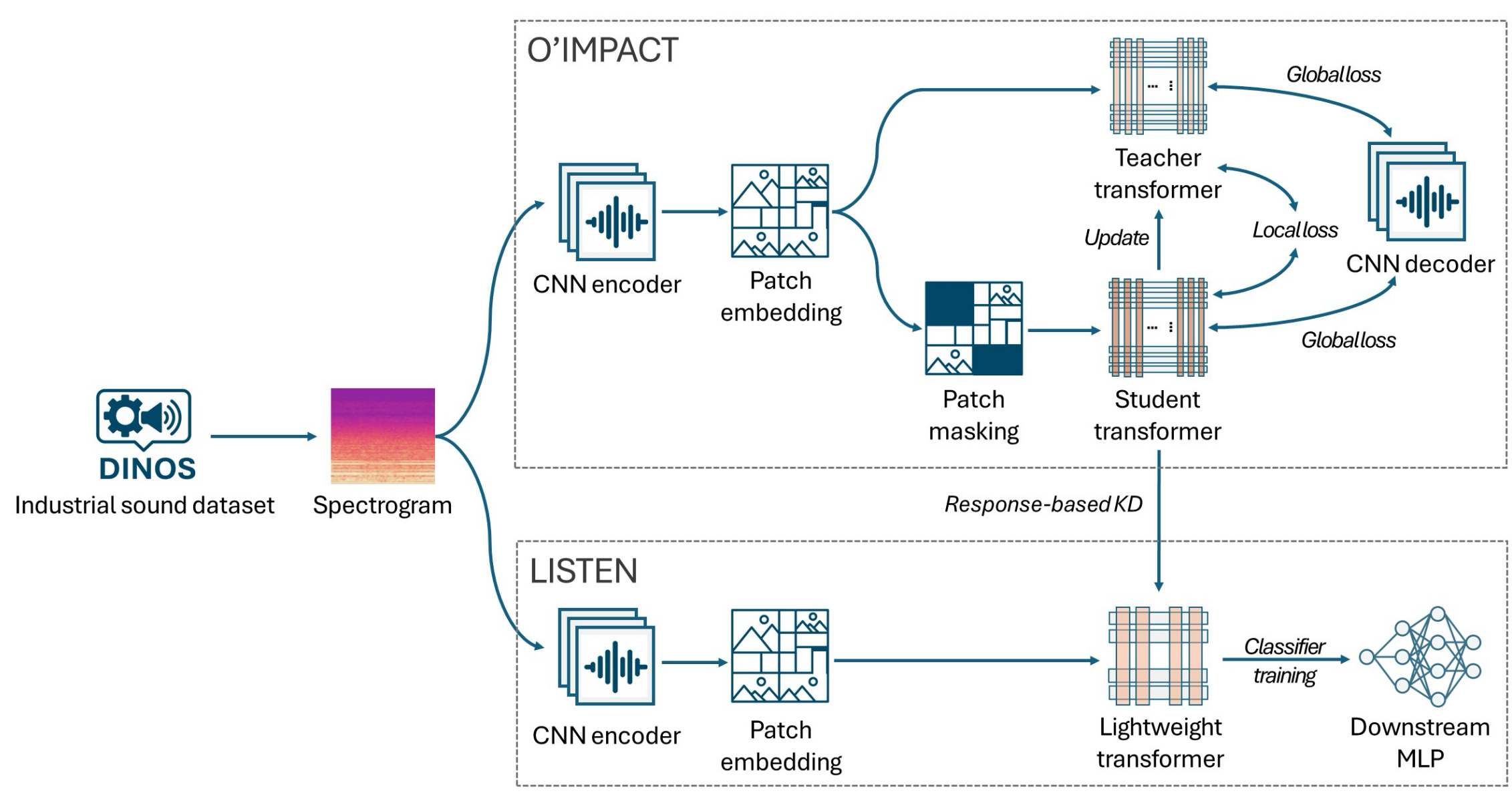

**Fig. 2** Pipeline of the Knowledge Distillation Process for LISTEN

**Table 1** Hyperparameters for the Grid Search on IMPACT

| Model name | I01 | I02 | I03 | I04 | I05 | I06 | I07 | I08 | I09 | I10 | I11 | I12 |
|---|---|---|---|---|---|---|---|---|---|---|---|---|
| Embedding dimensions | 128 | 128 | 128 | 192 | 192 | 192 | 256 | 256 | 256 | 384 | 384 | 384 |
| Number of layers | 4 | 6 | 8 | 4 | 6 | 8 | 4 | 6 | 8 | 4 | 6 | 8 |

The IMPACT model is a self-supervised learning framework built upon the EAT architecture [13]. This architecture employs a student-teacher training approach, comprising two identical transformer models. During training, the student model processes spectrograms with a 70% masking, while the teacher model is fed the complete spectrogram. The weights of the teacher model are not updated via backpropagation but by an Exponential Moving Average (EMA) strategy [43] each training epoch, a mechanism that enables the student to continuously learn from the data.

The training process is driven by a dual-objective loss function designed to capture features at different granularities. The first objective, the global loss ($L_{global}$), is calculated between the student's CLS token output and a global average-pooled representation of the teacher's final layer output. This forces the student model to learn a compressed, high-level summary of the entire spectrogram, mirroring the teacher's global understanding. The second objective, the local loss ($L_{local}$), is computed by passing the patch-level outputs of both models through a shared CNN decoder and then measuring the error between the reconstructed spectrograms. This ensures the student learns to reconstruct the fine-grained, local details and textures within the spectrogram.

These two losses are jointly optimized using the following total loss function:

$$L_{total} = L_{local} + \lambda L_{global} \quad (1)$$

where $\lambda$ is a coefficient that balances their relative weights. Industrial sounds are characterized by stable tonal harmonics associated with machine kinematics, broadband noise from physical processes, and distinctive temporal patterns such as periodic operations or fault-related transients [10,44,45]. Monitoring such monotonous and repetitive signals necessitates distinguishing subtle changes in fine-grained features. As shown in Table 2, the F1-scores drop drastically when $\lambda$ is set to 0.6. A high $\lambda$ biases the model toward global representations, causing it to neglect critical local features in the signal. Consequently, we selected $\lambda = 0.1$ to optimally balance an accurate global summary (global loss) with precise local reconstruction (local loss), enabling IMPACT to learn a comprehensive feature set suitable for diverse downstream tasks.

**Table 2** Impact of the Loss Weight Balancing Coefficient ($\boldsymbol{\lambda}$) on Downstream Task F1-scores

| | $\lambda = 0.0$ | $\lambda = 0.05$ | $\lambda = 0.1$ | $\lambda = 0.3$ | $\lambda = 0.6$ | $\lambda = 0.8$ |
|---|---|---|---|---|---|---|
| **ColdSpray** | 0.897 | 0.897 | 0.919 | 0.910 | 0.676 | 0.859 |
| **Renishaw** | 0.999 | 1.000 | 1.000 | 1.000 | 0.990 | 0.998 |
| **Yornew** | 0.878 | 0.881 | 0.900 | 0.878 | 0.990 | 0.874 |
| **VF2** | 0.935 | 0.943 | 0.952 | 0.947 | 0.571 | 0.935 |

Both losses are quantified using the Huber loss. We chose this loss function due to its robustness to outliers [46], which are common in industrial acoustic data in the form of transient noise spikes. Unlike MSE, which heavily penalizes large errors, Huber loss uniquely combines the advantages of MSE for small errors and Mean Absolute Error (MAE) for large errors. This characteristic makes it less sensitive to extreme outliers while still providing stable gradients for smaller variations, leading to a more robust training process.

LISTEN comprises three sequential components: a CNN encoder, a patch embedding layer, and lightweight transformers. First, the CNN encoder captures local spectral–temporal correlations in the Mel-spectrogram, effectively modeling fine-grained structures such as neighboring frequency interactions and short-term temporal variations. Next, the patch embedding layer partitions the feature map into non-overlapping patches and projects them into a compact token sequence. Finally, the lightweight transformers process this sequence using self-attention to model inter-patch relationships, capturing both fine-grained local features and global dependencies such as long-range temporal patterns and periodic structures characteristic of industrial machine operations.

A separate grid search was conducted to determine the optimal configuration of LISTEN. Table 3 lists the hyperparameters evaluated during this search. With the exception of the activation function, all other transformer and CNN encoder hyperparameters were kept identical to those of the O'IMPACT model. To reduce the computational load for an edge device, LISTEN employs the ReLU activation function instead of the GELU function used by O'IMPACT. This choice was a deliberate trade-off that prioritized architectural simplicity and computational efficiency. While the GELU function can be effective, its computational complexity increases due to its probabilistic nature [47,48]. In contrast, the simple comparison operation of ReLU offers a more lightweight and efficient alternative, making it a more pragmatic choice for our resource-constrained target hardware.

LISTEN was subsequently trained via KD. Distillation strategies are broadly categorized into response-based, feature-based, and relation-based methods [49]. Feature-based methods require matching intermediate feature maps, which constrains the student's architecture and complicates the training process. Relation-based methods add another layer of complexity by exploring inter-sample relationships. In contrast, we chose a response-based approach, framing the distillation task as a direct regression problem. We therefore minimized the MSE between the teacher's and student's final outputs, which compels the student model to learn and predict the teacher's raw logit values [50].

**Table 3** Hyperparameters for the Grid Search on LISTEN

| **Model name** | **L01** | **L02** | **L03** | **L04** | **L05** | **L06** | **L07** | **L08** | **L09** |
|---|---|---|---|---|---|---|---|---|---|
| **Embedding dimensions** | 16 | 16 | 16 | 16 | 16 | 16 | 16 | 16 | 16 |
| **Number of layers** | 2 | 2 | 2 | 4 | 4 | 4 | 6 | 6 | 6 |
| **Expansion factor** | 1 | 2 | 4 | 1 | 2 | 4 | 1 | 2 | 4 |
| **Model name** | **L10** | **L11** | **L12** | **L13** | **L14** | **L15** | **L16** | **L17** | **L18** |
| **Embedding dimensions** | 32 | 32 | 32 | 32 | 32 | 32 | 32 | 32 | 32 |
| **Number of layers** | 2 | 2 | 2 | 4 | 4 | 4 | 6 | 6 | 6 |
| **Expansion factor** | 1 | 2 | 4 | 1 | 2 | 4 | 1 | 2 | 4 |
| **Model name** | **L19** | **L20** | **L21** | **L22** | **L23** | **L24** | **L25** | **L26** | **L27** |
| **Embedding dimensions** | 64 | 64 | 64 | 64 | 64 | 64 | 64 | 64 | 64 |
| **Number of layers** | 2 | 2 | 2 | 4 | 4 | 4 | 6 | 6 | 6 |
| **Expansion factor** | 1 | 2 | 4 | 1 | 2 | 4 | 1 | 2 | 4 |

MSE is the most direct loss function for such a regression task. This approach compels the student to learn not just the teacher's predictions, but also its internal representational geometry, which can be a more powerful form of knowledge transfer. Furthermore, it avoids the need to tune an additional hyperparameter like temperature, which is required for cross-entropy-loss-based KD, thus aligning with our goal of a simple and robust training pipeline. This method provides greater architectural flexibility and computational efficiency. It directly focuses the student on mimicking the teacher's final inference, making it highly suitable for our goal of aggressive and independent lightweight model design.

The scalability and generalization of our approach were validated on the DINOS benchmark of 30 distinct downstream tasks spanning four different types of industrial machinery. The difficulty of these tasks was designed to be diverse, ranging from distinguishing basic states like on/off to classifying more complex, realistic scenarios such as chatter states or multiple distinct fault types. We freeze the backbones and train a shallow MLP head on downstream data to evaluate models' scalability and generalizability to unseen physical equipment, machine models, and processes, without full fine-tuning or retraining on a large-scale dataset. To ensure a rigorous evaluation across this benchmark, we employed a repeated random sub-sampling validation protocol. In each of the independent 10 experimental runs per task, the data was randomly partitioned into a 20% training set and a strictly separate 80% hold-out test set. The model was trained exclusively on the small training partition and evaluated on the large, unseen test set. This unconventional split was designed to simulate a real-world scenario where only a small amount of labeled data is available for model adaptation, thereby rigorously testing the model's data efficiency.

All models were developed using the DINOS dataset, which contains one-second industrial machine sound clips. Each clip was transformed into a log-Mel spectrogram of dimension $1 \times 128 \times 128$ using the following parameters: a 2,048-point Fast Fourier Transform (FFT), a 2,048-sample window length, a 376-sample hop length, 128 Mel bands, and a top decibel level of 80. These spectrograms were first processed by a CNN encoder, then divided into $16 \times 16$ non-overlapping patches and embedded. Table 4 summarizes the downstream MLP settings. All preprocessing, training, and postprocessing tasks were conducted in a WSL2 Ubuntu 22.04.5 LTS environment using PyTorch 2.7.0. The system was equipped with an Intel Core i9-14900HX GPU, 32 GB of RAM, and an NVIDIA GeForce RTX 4090 Laptop GPU.

**Table 4** Configuration of the Downstream MLP

| Type | Input shape | Output shape | Type | Input shape | Output shape | Type | Input shape | Output shape |
|---|---|---|---|---|---|---|---|---|
| **Layer 1 (ReLU)** | Embedding dimensions | 256 | **LayerNorm** | 256 | 256 | **Layer 2** | 256 | Number of classes |

## 3.2. On-site System Implementation and Validation

### 3.2.1. General Methodology for On-site Deployment

This section details the methodology for our proposed end-to-end system for on-device machine monitoring. The workflow is designed to be adaptable to new industrial environments with minimal overhead. The process begins by installing a sound sensor to a target machine to capture its acoustic signatures. For each distinct operational mode or task to be monitored, a small dataset is collected, often from just a single trial. This data is then transferred to a standard PC to adapt the pre-trained LISTEN model. Once adapted, the specialized model is deployed to a low-cost edge device. In the final stage, the edge device performs real-time inference on the live audio data streamed directly from the machine's sensor.

### 3.2.2. Case Study: Real-time Monitoring of a Live CNC Machine

To provide an initial validation of this methodology, we implemented the system in a real-time monitoring scenario on a Yornew VMC300 CNC machine. The experiment involved machining an AL6061 aluminum block with a 2-flute high-speed steel end mill (YG-1, 01047, 1/4-inch diameter). The ten operational modes defined for this study are listed in Table 5. Throughout all experiments, the machine's feed rate and radial depth of cut were fixed at 2 mm/s and 6 mm, respectively.

As illustrated in Fig. 1 (e), the system hardware comprised a custom stethoscopic sound sensor built from a stethoscope (MDF Instruments Dual Head) and a microphone (Fifine K053), a Raspberry Pi 4, and a laptop. The stethoscopic sound sensor was chosen to effectively prevent high-frequency ambient noise while better isolating the machine's localized sounds [51,52]. Notably, this sensor does not require placement immediately adjacent to the sound source and was attached to the bottom of the machine's workbench for the experiment. The sensor was connected via a USB interface to a Raspberry Pi 4, which served as a low-cost, headless edge device for both data acquisition and inference.

Following our methodology—shallow head training on minimal data with a frozen backbone, we adapted the LISTEN model using data from a single execution per mode, capturing around 20 seconds of audio for each of the ten modes. This model adaptation process was conducted on the same system used for the initial KD phase. The specialized model was then deployed on a Raspberry Pi 4, which served as a low-cost, headless edge device. The Raspberry Pi performed real-time inference on sequential one-second audio clips streamed from the sensor, transmitting the results to a remote PC via a wireless network for monitoring.

**Table 5** Modes of the CNC Machining Processes

| Mode name | Mode 0 | Mode 1 | Mode 2 | Mode 3 | Mode 4 | Mode 5 | Mode 6 | Mode 7 | Mode 8 | Mode 9 |
|---|---|---|---|---|---|---|---|---|---|---|
| **Axial depth** | Off | On | 1 mm | 1 mm | 1 mm | 1 mm | 3 mm | 3 mm | 3 mm | 3 mm |
| **Spindle speed** | Off | On | 6000 rpm | 8000 rpm | 10000 rpm | 12000 rpm | 6000 rpm | 8000 rpm | 10000 rpm | 12000 rpm |

# 4. Results and Discussion

## 4.1. Selection of the Optimal Teacher Model

Fig. 3 presents a detailed performance analysis of the IMPACT architecture across various hyperparameter configurations to identify the optimal teacher model, with comprehensive results available in Appendix Fig. A1. For clarity, the analysis is divided into two plots. Plot (a) displays the overall average F1-score across all 30 downstream tasks, providing a measure of general performance. Plot (b) specifically details the F1-score on the unseen domain (ColdSpray) which serves as a key measure of generalization to new data. In both plots, the two final candidates, I05 and I09, are highlighted in red and blue, respectively, for direct comparison.

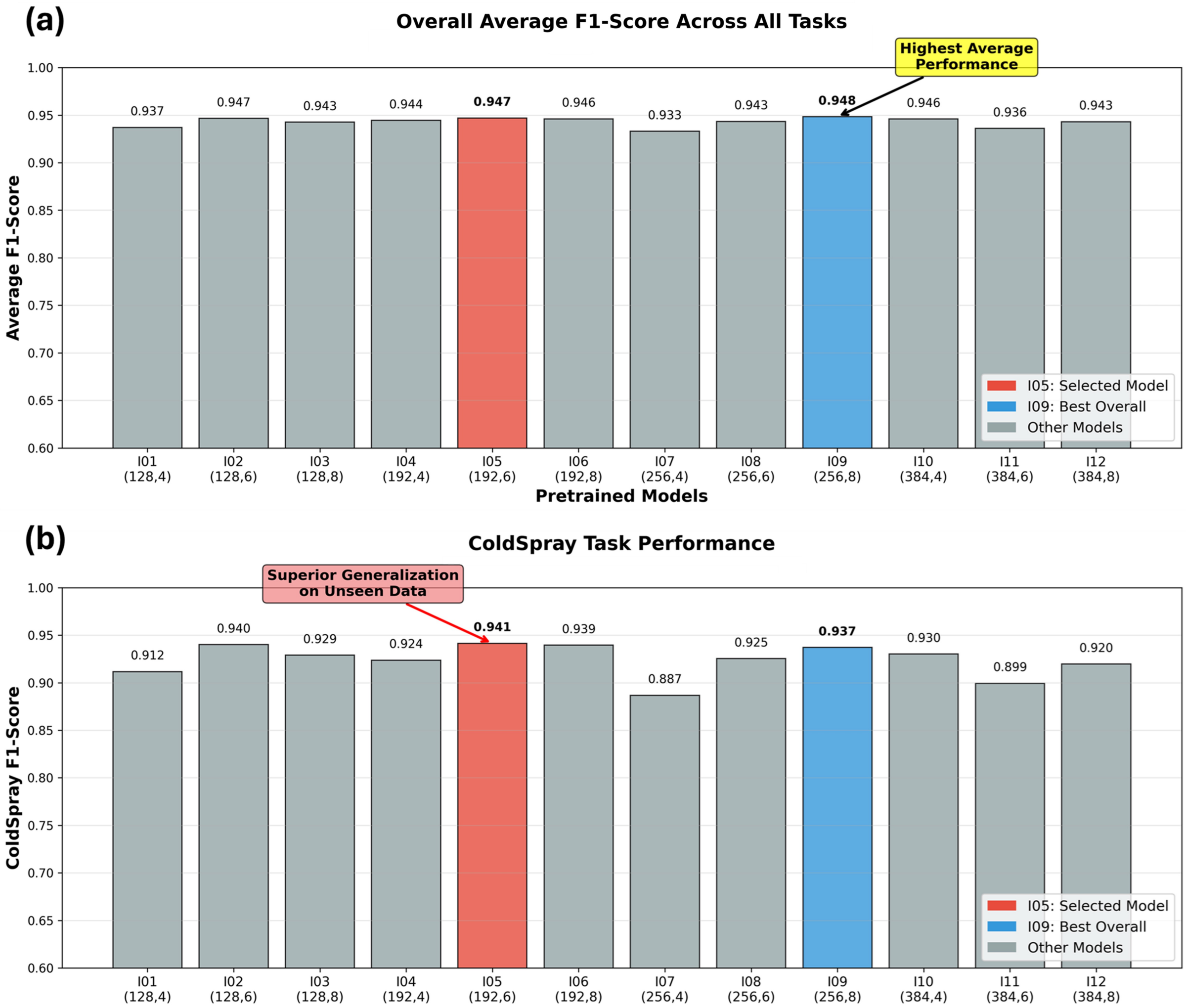

**Fig. 3** Performance Analysis for the O'IMPACT Hyperparameter Selection, (a) Overall Average F1-score, (b) Generalization Performance on the Unseen ColdSpray Dataset

The plots reveal the trade-off that informed our selection. While the I09 configuration achieved a marginally higher overall F1-score in plot (a), the performance difference was only 0.1%. In contrast, plot (b) shows that the I05 model's F1-score was 0.4% higher on the critical ColdSpray generalization task. Given that robust generalization to new, unseen data is a primary requirement for a foundation model, we selected the I05 model as the optimal parent (O'IMPACT). The selected O'IMPACT model contains approximately 4.23M parameters, with a total model size of 16.2 MB. When deployed on a Raspberry Pi 4, the model's inference time ranged from 64.9 ms to 180.5 ms per sample, failing to achieve real-time performance at 30 FPS (33.3 ms) [53].

## 4.2. LISTEN: Lightweight Model Selection via Performance-Efficiency Trade-off

LISTEN employs an ultra-lightweight architecture comprising a $16 \times 16$ strided convolutional patch embedding to aggressively compress an input into 64 tokens and a compact two-layer Transformer encoder with 64 embedding dimensions, designed for extreme inference efficiency on edge devices.

Using O'IMPACT as a teacher model, we conducted a grid search to identify the optimal configuration for the LISTEN model. Fig. 4 presents the results, visualizing the trade-off between model performance (F1-score) and efficiency (inference time). Fig. 4 presents the results of this search, visualizing the trade-off between model performance (F1-score) and efficiency (inference time). Comprehensive results for all configurations are provided in Appendix Fig. A2. Each point on the plot represents a unique hyperparameter configuration, with the most desirable models located in the top-left corner, offering high performance with low computational cost.

The plot clearly illustrates the rationale for our final model selection. The L22 configuration (highlighted in blue) achieved the highest overall F1-score, but at a significant efficiency cost. In contrast, the L19 configuration (highlighted in red) is positioned on the figure, offering a superior balance. While the performance gap between L19 and L22 was a marginal 0.1%, L19 required only 75.0% of the inference time and had half the number of parameters (50K vs. 100K) in the transformer layers. This compelling trade-off makes L19 the definitive choice for a model designed for efficient, real-world edge deployment.

This knowledge distillation process successfully reduced the model parameters from O'IMPACT's 4.23M to 0.07M for the final L19-based LISTEN model. Notably, its overall mean inference time on a Raspberry Pi 4 was 31.7 ms, comfortably meeting the real-time performance criterion of 33.3 ms (30 FPS). Additionally, we distilled O'IMPACT using popular lightweight architectures, such as MobileNetV4- S [54], MobileViT-XXS [29], and BC-ResNet-3 [55]. Table 6 details the number of parameters, Multiply–Accumulate Operations (MACs), inference time per sample, resource usage, and F1-score for each model.

**Table 6** Comparison of Performance and Inference Time between Distilled Models with Various Lightweight Architectures (Parameters and MACs: Encoder only)

| Model | Parameters | MACs | Time per sample | CPU/ Memory usage | F1-Score | | | |
|---|---|---|---|---|---|---|---|---|
| | | | | | ColdSpray | Renishaw | Yornew | VF2 |
| IMPACT | 17.3 M | 291.3 M | 179 ms | 86% / 454 MB | 0.919 | 1.000 | 0.900 | 0.952 |
| O'IMPACT | 4.23 M | 71.3 M | 92 ms | 77% / 394 MB | 0.941 | 1.000 | 0.891 | 0.955 |
| MobileNetV4-S | 1.45 M | 60.32 M | 78 ms | 82% / 386 MB | 0.909 | 1.000 | 0.906 | 0.957 |
| MobileViT-XXS | 0.76 M | 65.67 M | 190 ms | 61% / 402 MB | 0.936 | 1.000 | 0.900 | 0.949 |
| BC-ResNet-3 | 0.08 M | 47.06 M | 149 ms | 87% / 406 MB | 0.918 | 1.000 | 0.876 | 0.946 |
| **LISTEN** | **0.07 M** | **5.24 M** | **32 ms** | **66% / 364 MB** | **0.934** | **1.000** | **0.907** | **0.956** |

LISTEN achieved superior overall performance with the fastest inference time. Although BC-ResNet-3 has the smallest number of parameters and low theoretical computational complexity, it did not outperform the CNN-based MobileNetV4-S in terms of inference speed. BC-ResNet employs broadcasting mechanisms to reduce parameter count and MACs. However, such operations may introduce additional data movement and less regular memory access patterns compared to standard convolutional operations. The relatively high memory usage observed in BC-ResNet-3 is consistent with this behavior, demonstrating that reductions in MACs do not necessarily translate into latency improvements on CPU-based edge devices.

Meanwhile, MobileViT-XXS achieved an F1-score comparable to LISTEN, demonstrating that Transformer-based architectures are effective at capturing the characteristics of industrial acoustic signals. Nevertheless, its inference speed on edge devices lags behind CNN-based models. The relatively low CPU utilization (approximately 60%) observed in MobileViT-XXS suggests the presence of inefficiencies beyond pure computation, potentially related to memory access and data movement. While standard CNNs typically benefit from compute-dense operations with regular and continuous memory access, the self-attention mechanism in Transformers often incurs additional latency due to memory-intensive operations, intermediate tensor generation, and data reshaping [29,54].

LISTEN addresses these limitations by adopting a hybrid strategy. By employing a single convolutional layer for patch embedding, the input is compressed into a compact sequence of 64 tokens. This reduced representation is then processed by an aggressively simplified Transformer encoder, mitigating memory overhead and computational burden. As a result, LISTEN successfully combines the representational strength of Transformers with the execution efficiency of convolutional models, enabling real-time inference on CPU-based edge devices.

## 4.3. On-site System Validation: A CNC Machining Case Study

For the model adaptation process—shallow head training on minimal data with a frozen backbone, we curated 20 seconds of sound data for each operational mode. This rapid adaptation took only 61 seconds to complete 200 epochs. The final specialized model, comprising the frozen LISTEN backbone and a trained MLP head, was then deployed on a Raspberry Pi 4. During the live test, one-second audio clips were streamed from the operating CNC machine to the Raspberry Pi, which then inferred the machine's current operational mode in real-time.

As shown in Fig. 5, the on-site LISTEN implementation achieved an overall F1-score of 0.938. While Mode 4 had the lowest score at 0.893, this was still higher than the baseline F1-score (0.891) of the original IMPACT model on Yornew. To evaluate efficiency, we measured the end-to-end processing time for each audio clip, which included pre-processing, inference, and post-processing. The average end-to-end processing time for all modes remained consistently below the 33.3 ms (30 FPS) threshold. These findings successfully confirm the feasibility of our proposed on-site deployment workflow, demonstrating that LISTEN can be rapidly adapted to effectively monitor a new industrial scenario in real-time using minimal data.

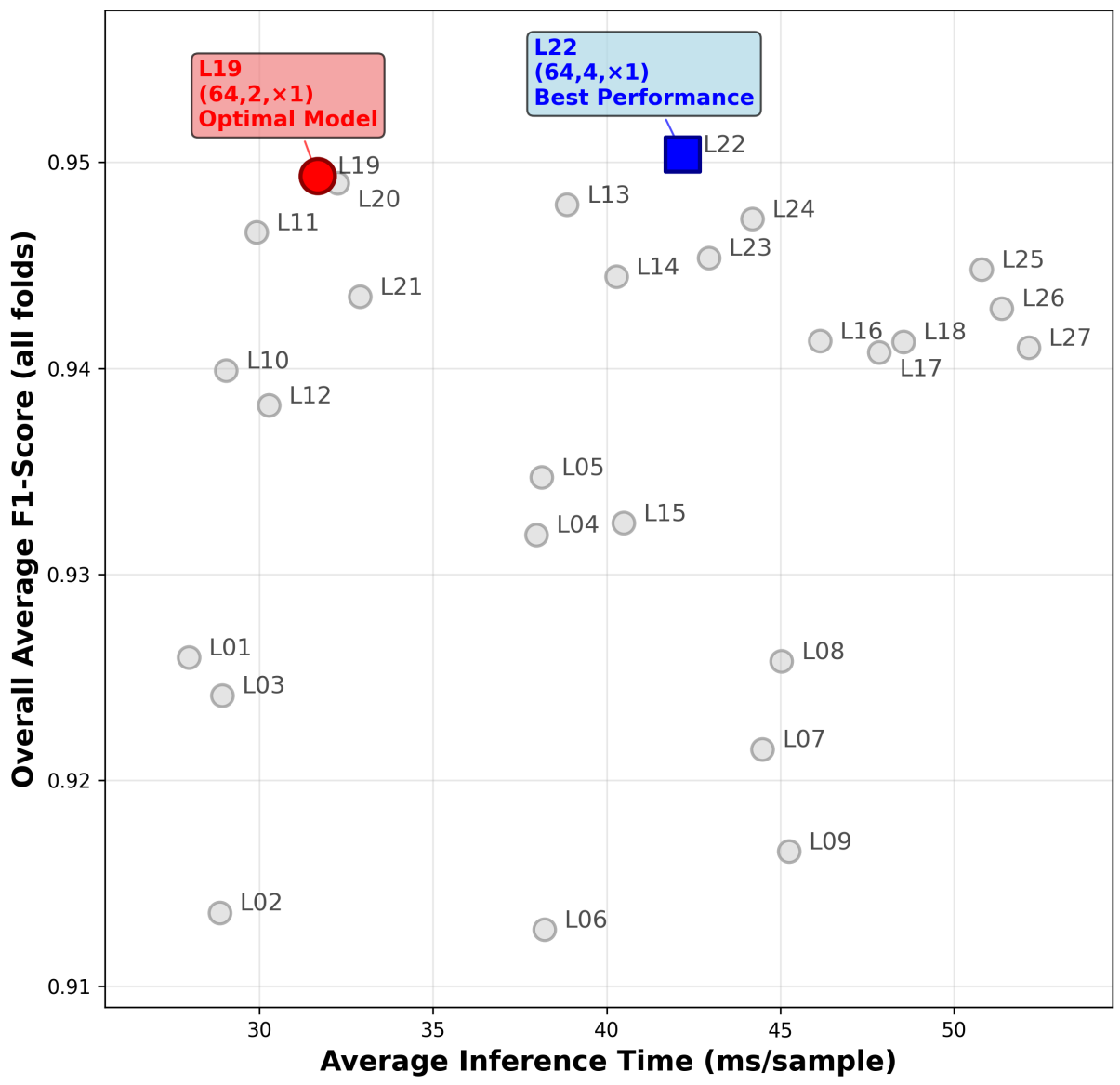

**Fig. 4** LISTEN Performance and Inference Time by Hyperparameter Configuration

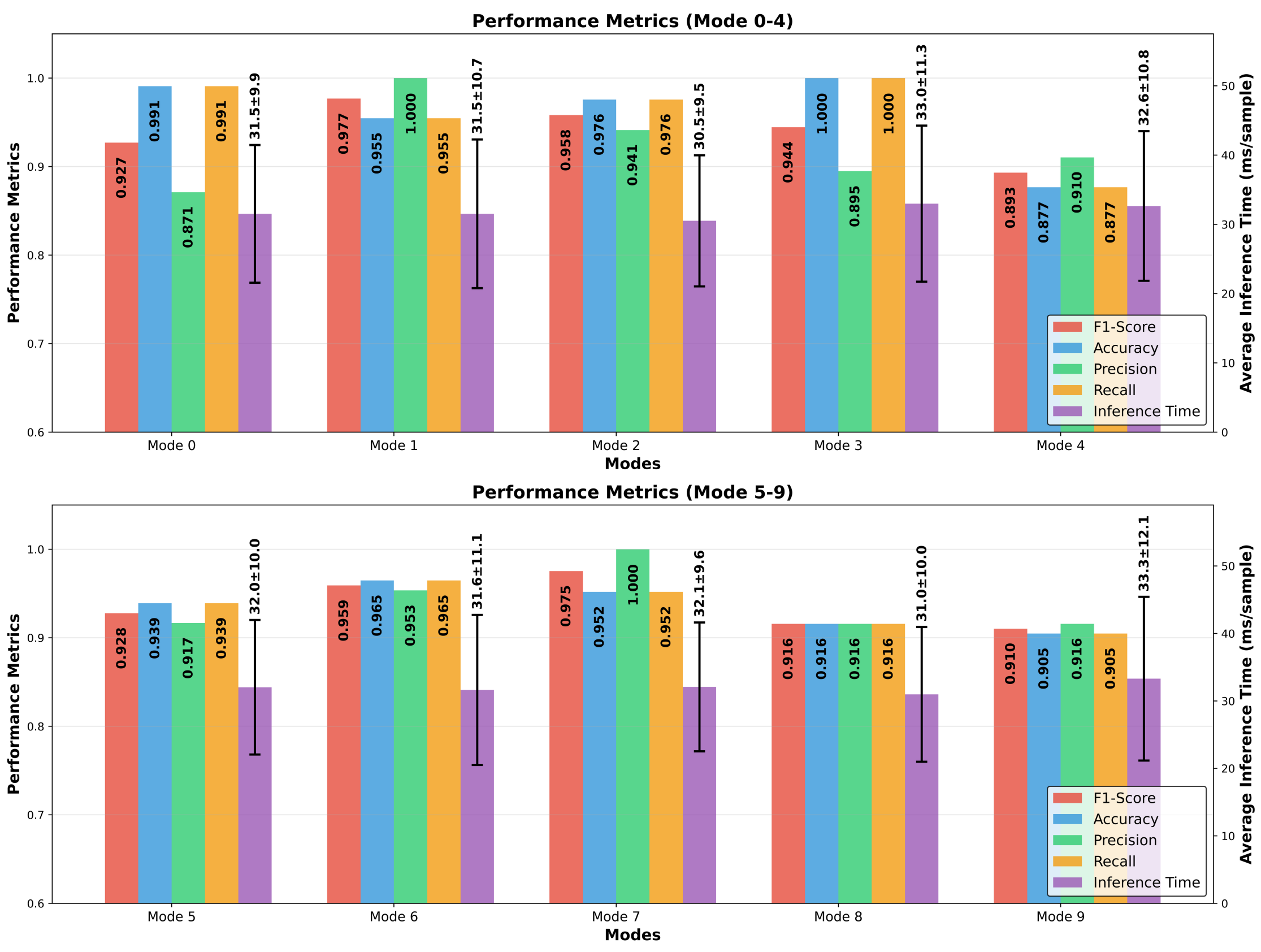

**Fig. 5** On-site System Validation Results

# 5. Conclusion

This study successfully bridged the critical gap between high-performance, domain-specific sound models and the practical constraints of on-device industrial monitoring. We introduced the first kilobyte-sized foundation model specialized for industrial sound and demonstrated a complete end-to-end system for its deployment. Through grid search, we identified an optimal lightweight architecture for LISTEN that retains the performance of the teacher model across diverse benchmarks. As validated in our live CNC scenario, LISTEN generalizes to new environments without structural re-optimization, requiring only the retraining of a shallow classifier using single-trial data.

Our workflow enables rapid on-site model adaptation in just 61 seconds with minimal data, drastically reducing the setup time and data dependency. Unlike traditional approaches that require full fine-tuning or training of the entire architecture, LISTEN enables rapid on-site adaptation requiring only a single trial of data and one minute of training for a shallow MLP head while keeping the backbone frozen. It further demonstrates strong scalability and generalizability across unseen physical equipment, machine models, and processes on the DINOS benchmark. By achieving real-time inference on a low-cost edge device while meeting performance baselines, we have established the practical viability of deploying advanced AI-driven machine listening system on live shop floors. Furthermore, the edge-native design of our framework inherently enhances cybersecurity by processing sensitive operational data locally.

LISTEN drastically reduces the complexity of deep learning-based machine listening on a shop floor while maintaining robustness to diverse environments. However, this study also revealed several limitations, presenting clear avenues for future work.

First, most on-site inference errors occur during transitional states, such as the spindle accelerating or at the beginning and end of a machining process. While we used a one-second audio clip length, future work should conduct a parametric investigation to determine the optimal length for improving LISTEN's practical resolution. Second, although LISTEN achieved real-time processing speeds, the performance margin was very small, and its sample-by-sample inference time often exceeded the 33.3 ms threshold. To create an even lighter and faster model, we plan to apply advanced techniques like quantization, which represent the model’s weights, biases, and activations in simpler formats. We are currently optimizing the O’IMPACT and LISTEN configurations to make them suitable for various quantization methods.

Third, although LISTEN exceeded the baseline, its performance does not yet meet the rigorous standards required for a commercial real-world monitoring system. Enhancing its performance requires a larger volume of data, which is often difficult to acquire readily in industrial settings. To address this, we are developing a data augmentation framework based on a physical sound propagation model, which will help generate synthetic sound data from actual source signals. Finally, LISTEN must be validated across more diverse scenarios. We plan to implement and test LISTEN in various industrial environments, including additive manufacturing, legacy machine operations, and different types of CNC machining.

Addressing these limitations is crucial for transitioning LISTEN from a validated prototype to a practical, robust system. We anticipate that these concerted efforts to optimize temporal resolution, enhancing computational efficiency through quantization, expanding the training data via physics-based augmentation, and validating across diverse scenarios will collectively advance its practical use beyond lab-scale applications. Ultimately, LISTEN establishes a clear and feasible pathway for deploying sophisticated, domain-specific AI on low-cost edge hardware. We envision that its continued evolution will enable truly autonomous, secure, and intelligent machine monitoring, transforming industrial acoustic data from a passive signal into an active resource for process optimization and predictive maintenance.

# Appendix

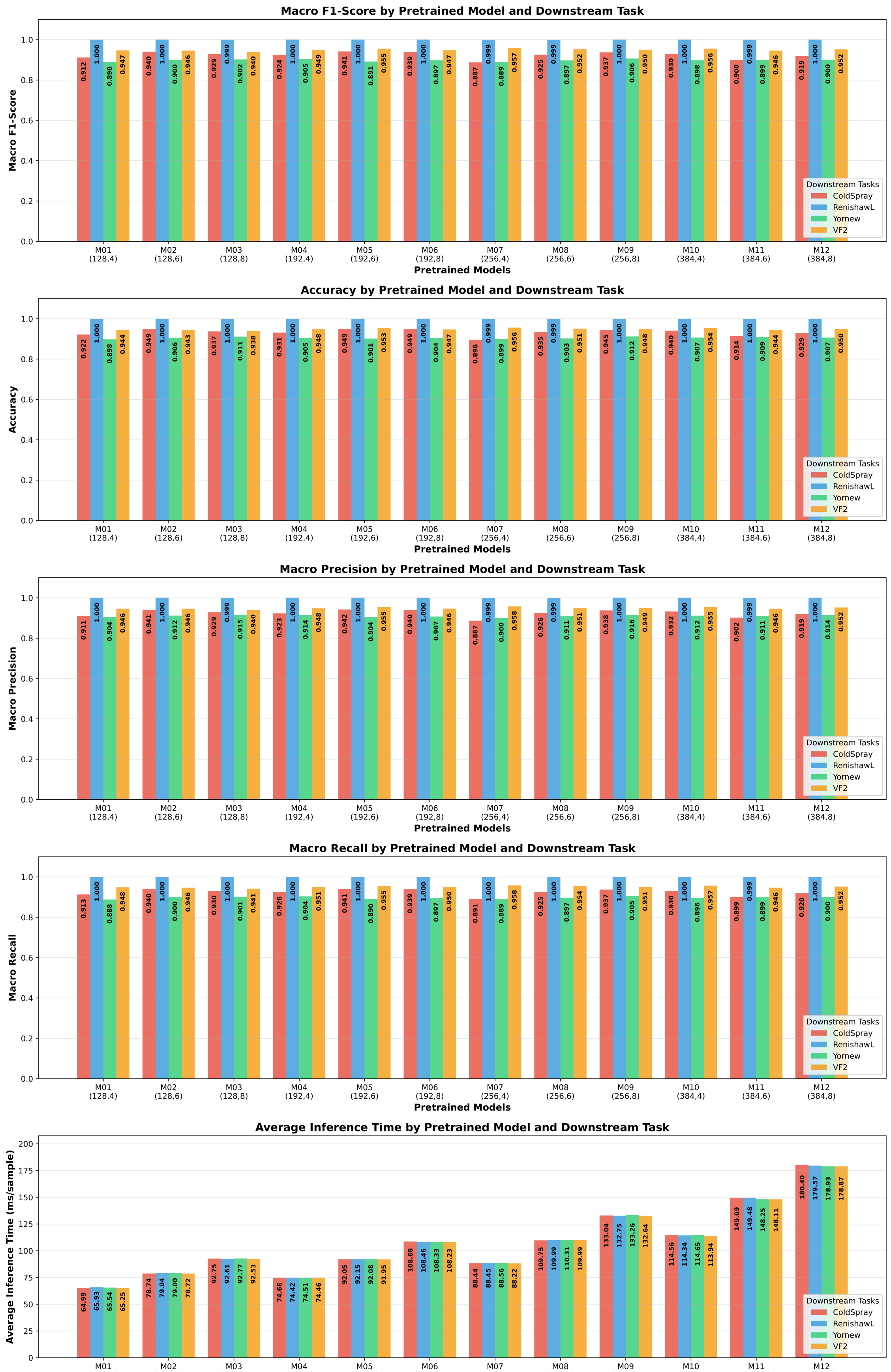

**Fig. A1** Comprehensive IMPACT Performance and Inference Time by Hyperparameter Configuration

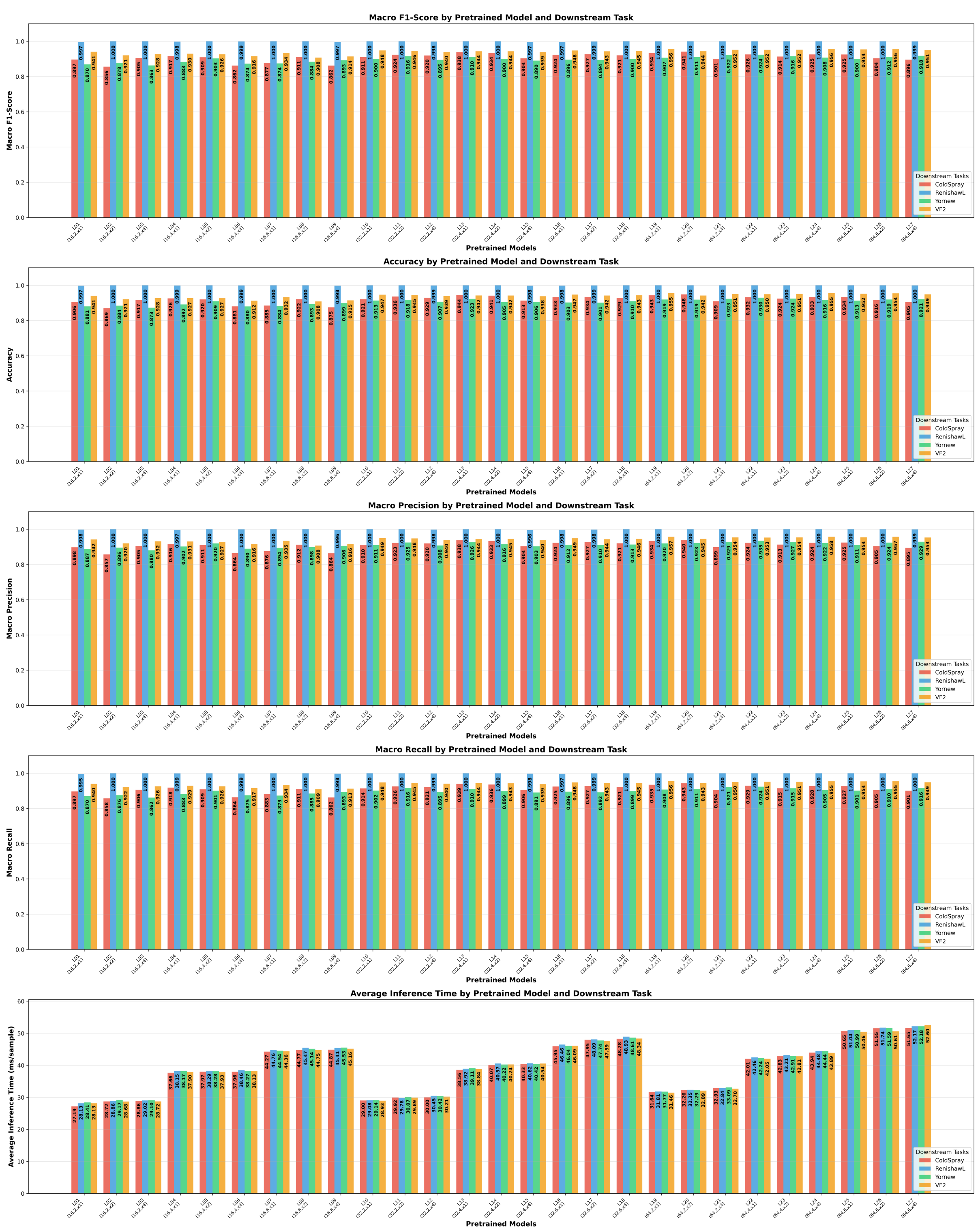

**Fig. A2** Comprehensive LISTEN Performance by Hyperparameter Configuration